\documentclass[floatfix,aps,prd,abstract,twocolumn,10pt,nofootinbib,preprintnumbers,superscriptaddress,usenames,dvipsnames ]{revtex4}

\usepackage{amsgen,amsmath,amstext,amsbsy,amsopn,amssymb}
\usepackage{graphicx}
\usepackage[usenames]{color}
\usepackage{epsfig}
\usepackage{longtable}
\usepackage{bm} 
\usepackage{ulem}
\usepackage{hyperref}
\usepackage{fontenc}
\usepackage{inputenc}
\usepackage{xcolor}
\usepackage{varwidth}

\usepackage{braket}

\newcommand{\be}{\begin{equation}}
\newcommand{\ee}{\end{equation}}
\newcommand{\bea}{\begin{eqnarray}}
\newcommand{\eea}{\end{eqnarray}}

\def\k{{\bf k}}
\def\d{{\rm d}}

\begin{document}

\title{Constraints on long-lived, higher-spin particles from galaxy bispectrum}

\author{Azadeh Moradinezhad Dizgah}
\affiliation{Department of Physics, Harvard University, 17 Oxford Street, Cambridge, MA 02138, USA }
\email{amoradinejad@physics.harvard.edu}
\author{Gabriele Franciolini}
\affiliation{D\'epartement de Physique Th\'eorique and Centre for Astroparticle Physics (CAP),\\
Universit\'e de Gen\`eve, 24 quai E. Ansermet, CH-1211 Gen\`eve, Switzerland}
\author{Alex Kehagias}
\affiliation{Physics Division, National Technical University of Athens, 15780 Zografou Campus, Athens, Greece}
\author{Antonio Riotto}
\affiliation{D\'epartement de Physique Th\'eorique and Centre for Astroparticle Physics (CAP),\\
Universit\'e de Gen\`eve, 24 quai E. Ansermet, CH-1211 Gen\`eve, Switzerland}

\begin{abstract}
The presence of massive particles with spin during inflation induces distinct signatures on correlation functions of primordial curvature fluctuations. In particular, the bispectrum of primordial perturbations obtains an angular dependence determined by the spin of the particle, which can be used to set constraints on the presence of such particles. If these particles are long-lived on super-Hubble scales, as is the case for example for partially massless particles, their imprint on correlation functions of curvature perturbations would be unsuppressed. In this paper, we make a forecast for how well such angular dependence can be constrained by the upcoming EUCLID spectroscopic survey via the measurement of galaxy bispectrum.  
\end{abstract}

\maketitle

\section{ Introduction } 
Inflation \cite{Lyth:1998xn} is a successful theory in solving the problems of standard big bang theory, namely flatness and horizon problems. Additionally, it provides a mechanism for generating primordial fluctuations which are the seed of the observed anisotropies in the cosmic microwave background (CMB) as well as the large scale structure (LSS) of the universe. The simplest models of inflation with a single degree of freedom, i.e., inflaton, originating from the Bunch-Davies vacuum, predict a nearly Gaussian distribution of primordial fluctuations. High precision constraints on the level of primordial non-Gaussianity \cite{Bartolo:2004if}, will shed light on the field content and interactions between quantum fields during inflation, and hence enable us to distinguish between inflation models.

Even when inflation is driven by a single degree of freedom, the excitation of additional particles present during inflation can leave an imprint on correlation functions of primordial curvature perturbations $\zeta$. In particular, it generates primordial non-Gaussianity which can be used to constrain the characteristics of such particles. The signatures of extra scalar fields, have been extensively studied in the context of curvaton models (see \cite{Byrnes:2010em} for a review), in the case of light scalar fields, and in the context of quasi-single field models \cite{Chen:2009we,Chen:2009zp,Baumann:2011nk,Noumi:2012vr,Kehagias:2015jha,Dimastrogiovanni:2015pla}, for the case of massive scalar fields with masses of the order of Hubble parameter during inflation. A general treatment of the impact of massive particles with spin on the correlation functions of primordial curvature perturbations has been recently studied in \cite{Arkani-Hamed:2015bza, Lee:2016vti}. It was shown that for particles with $m\sim H$ during inflaton, the primordial bispectrum has an angular-dependence (depending on the angles between the three wavevectors) determined by the spin of the particles and an oscillatory (or power-law) feature determined by its mass.

The so-called Higuchi bound \cite{Higuchi:1986py}, implies that the massive spinning fields in de-Sitter background decay on super-Hubble scales; hence, they are short-lived and their imprint on the cosmological correlation functions is suppressed. There are two ways in which one can generate long-lived massive particles with spin, thus unsuppressed perturbations on super-Hubble scales. One is by introducing a suitable coupling between inflaton and the extra particle with spin \cite{Kehagias:2017cym,Franciolini:2017ktv}, in analogy to the previously studied case of vector fields coupled to inflaton \cite{Barnaby:2012tk,Bartolo:2012sd,Shiraishi:2013vja,Bartolo:2015dga}. Another possibility is the  partially massless particles \cite{Deser:1983tm,Deser:2001pe,Deser:2001us,Deser:2003gw} which,  for some  
discrete values of the mass of the particle with spin $s$ are characterized by long-lived perturbations on super-Hubble scales for certain helicity states. In both ways, and in the context of exact de Sitter (which amounts to assuming the curvature perturbation is generated not by the inflaton, but by a spectator field and therefore multi-field models of inflation) it was shown in Ref. \cite{Franciolini:2017ktv} that spinning field obtains  a non-zero vacuum expectation value which introduces preferred direction leading to the statistical anisotropy of the cosmological  correlators, in particular in the power spectrum, bispectrum, and trispectrum\footnote{Partially massless particles  were considered also in Ref. \cite{Baumann:2017jvh}, but within the single-field models of inflation and with no vacuum expectation value for the spinning fields, thus giving rise to a non-vanishing trispectrum of curvature perturbations while the bispectrum is vanishing.}.

\begin{figure*}[t]
  \begin{tabular}{c}
    \begin{minipage}{1.\hsize}
  \begin{center}
    \includegraphics[width = 0.7 \textwidth]{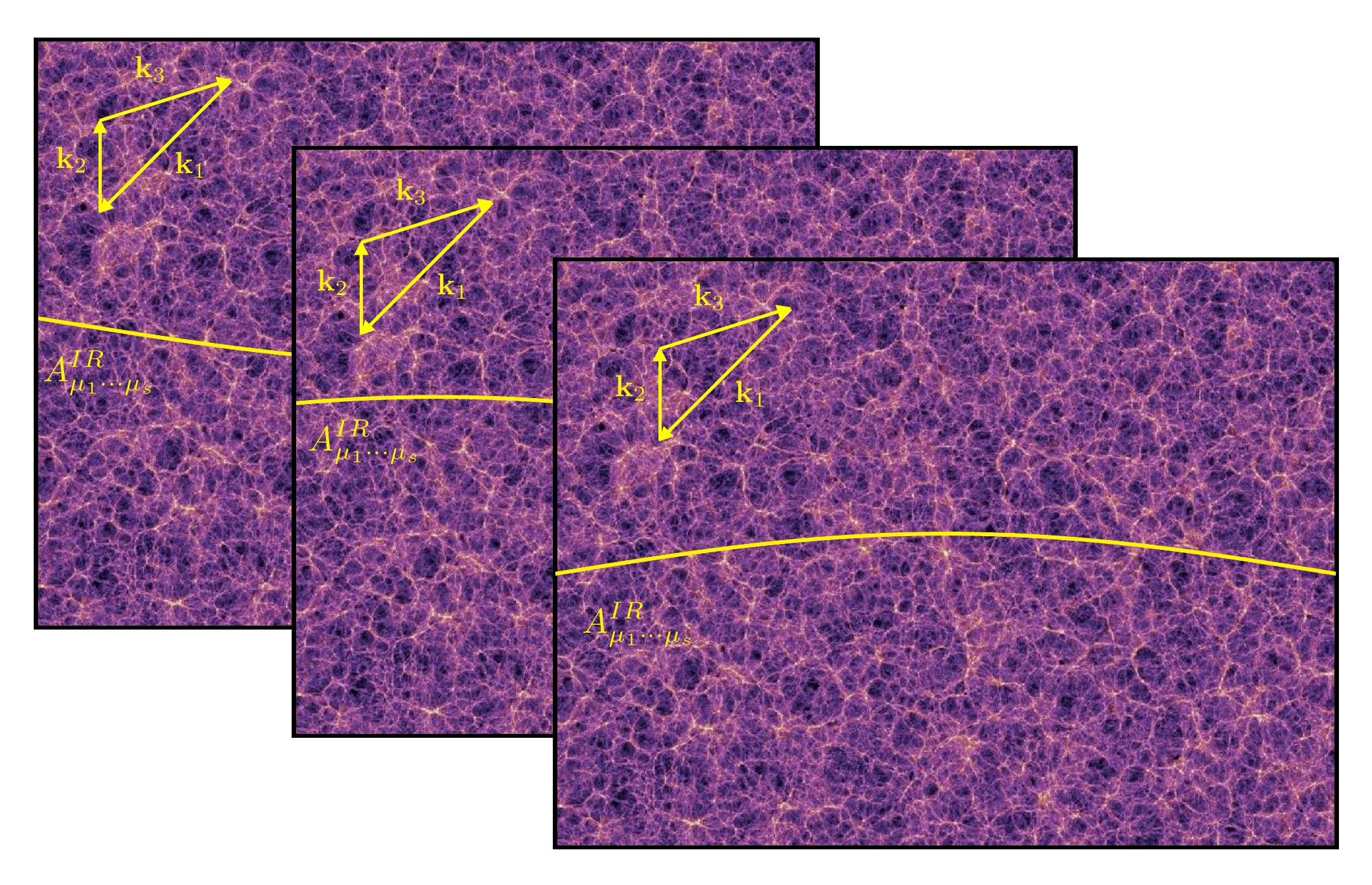}
  \end{center}
\end{minipage}
\end{tabular}
  \caption{A schematic representation of various realizations of the first $N-N_k$ e-folds of inflation in which the long-lived IR (i.e. super-Hubble) higher-spin modes $A_{\mu_1 \cdots \mu_s }^{IR}$ act as non-trivial background. The cosmological perturbations depend on the particular value the IR modes assume in a  \textsl{single} realization of the ensemble of possible universes.}\label{fig:euc} 
\end{figure*}

 In this paper, we investigate detectability of signature of such long-lived higher spin (HS) fields through their imprint on galaxy bispectrum.
Current best constraints on various shapes of primordial non-Gaussianity are from measurements of temperature and polarization bispectra of the CMB by the Planck satellite \cite{Ade:2015ava}. Further improvement of these constraints are expected to be achieved via analysis of clustering statistics of LSS from upcoming galaxy surveys such as DESI \cite{Font-Ribera:2013rwa}, EUCLID \cite{Amendola:2016saw}, and LSST \cite{Abell:2009aa}. Additionally, intensity mapping technique can also potentially be used as a tracer of large scale structure and hence provide a mean to constrain primordial non-Gaussianity.

There are several forecasts for constraints on primordial non-Gaussianity of local, equilateral, and orthogonal shapes from upcoming galaxy and intensity mapping surveys (see for ex. \cite{Giannantonio:2011ya,Tellarini:2016sgp,Yamauchi:2016wuc,Baldauf:2016sjb,Karagiannis:2018jdt,Camera:2013kpa,Munoz:2015eqa,Li:2017jnt,MoradinezhadDizgah:2018zrs}). The constraints on the presence of additional (short-lived) massive particles with and without spin from LSS were obtained in \cite{Sefusatti:2012ye,Meerburg:2016zdz,MoradinezhadDizgah:2017szk,MoradinezhadDizgah:2018ssw}. For the long-lived particles with spin, their detectability in CMB and galaxy power spectra, as well as the CMB bispectrum, is studied in \cite{Bartolo:2017sbu,Franciolini:2018eno}. We extend their analysis to obtain constraints from the galaxy bispectrum. 

The analysis of the imprint of the full anisotropic bispectrum on CMB and LSS bispectrum is rather complex, if not unfeasible.  One can instead consider the angle-average bispectrum as is done in Ref. \cite{Franciolini:2018eno} to search for the imprint of these long-lived higher-spin particles. It was shown in \cite{Franciolini:2018eno} that the angle-averaged bispectra due to massive and partially massless particles with a given spin, can be expanded in terms of Legendre polynomials with a finite number of terms. We use this result to constrain the coefficients of this expansion from measurements of galaxy bispectrum from upcoming EUCLID survey.

The rest of the paper is organized as follows: in section \ref{sec:theory}  we review the features of the bispectrum of curvature perturbations due to the presence of long-lived higher-spin fields and the template of primordial bispectrum that we use in our forecast. In section \ref{sec:gbis} we review our model of the observed galaxy bispectrum, while in section \ref{sec:forecast_meth}, we outline our forecasting methodology. We present our results in section \ref{sec:results} and conclude in section \ref{sec:conclusions}.

\section{Primordial bispectrum due to long-lived spinning particles}\label{sec:theory}
The impact of spinning particles on cosmological correlators can be analysed with the aid of the dS/CFT$_3$ correspondence \cite{Strominger:2001pn}, which can be used under the assumption that inflation is realized as a phase of (quasi-)de Sitter spacetime in the early universe. On the boundary of dS spacetime, i.e., in the limit where the conformal time $\tau$ tends to zero, the HS field can be written as $A_{i_1 \cdots i_s}(\vec{x} , \tau) = (-\tau )^{\Delta-s}A_{i_1 \cdots i_s}(\vec{x})$ where $\Delta= {3}/{2} - \sqrt{(s-{1}/{2})^2-{m^2}/{H^2}}$. One can easily notice that the value of conformal weight $\Delta = 0$ is indeed a very special case, as it is the only one giving rise to healthy long-lived perturbations. For values $\Delta < 0$ one encounters unphysical divergences of the ultraviolet modes stretching over the horizon scale during inflation. On the other hand, for positive values of the conformal weight, the characteristic amplitude of the HS fluctuation  rapidly decays on super Hubble scales. In particular, the bispectrum contribution from the spinning particles in the squeezed limit configuration is suppressed by a factor $(k_{\rm long}/k_{ \rm short})^{\Delta}$ \cite{Arkani-Hamed:2015bza}.
While the Higuchi bound \cite{Higuchi:1986py} dictates that the conformal weight $\Delta$ must be higher than one due to unitarity bounds, there are at least two ways in which  $\Delta=0$ can be achieved, namely through an ``ad-hoc" coupling between the inflaton and the HS fields or by restricting the analysis to partially massless HS fields, as stated in the introduction.

In order to parametrize the impact of spinning particles on the bispectrum we adopt the following template
\be \label{eq:temp}
B_\zeta(k_1,k_2,k_3) = \sum_{n} C_n {\mathcal P}_n(\hat k_1\cdot\hat k_2)  P_\zeta(k_1)P_\zeta(k_2) + 2\ {\rm perms},
\ee
where ${\mathcal P}_n$ are Legendre polynomials of order $n$ and $P_\zeta(k)$ is the power spectrum of primordial fluctuations. As we will review below, the long-lived particles with spin $s$ induce $s+1$ non-vanishing coefficients with n even: $C_0, C_2, \dots , C_{2s-2}, C_{2s}$. 

Referring to \cite{Franciolini:2017ktv,Franciolini:2018eno} for details, we present only the main results relevant for the present discussion. In the case of HS fields coupled to the inflaton it was found that 
 the statistically anisotropic curvature bispectrum takes the form \cite{Franciolini:2017ktv}:
\begin{eqnarray} \label{eq:HS_bis_gen}
B_\zeta(\k_1,\k_2,\k_3) 
  &=& g_s \Braket{ \bar{A}^{i_1\cdots i_s}} \Braket{ \bar{A}^{j_1\cdots j_s}} P_\zeta(k_1)P_\zeta(k_2) \nonumber \\ 
&&\hspace{-20pt}  \times \ \Pi_{i_1\cdots i_s}^{\ell_1 \cdots \ell_s} (\k_1)\Pi_{j_1\cdots j_s}^{\ell_1 \cdots \ell_s} (\k_2)  +  2~{\rm perms},  \label{eq:bis-hs}
\end{eqnarray} 
where $g_s$ is an undetermined constant proportional to the spin-dependent coupling between the HS field and the inflaton, and the projector tensor $\Pi_{i_1\cdots i_s}^{j_1 \cdots j_s} (\k)$ is built as the sum of helicities of the HS polarization tensors
\be
\Pi_{ i_1 \cdots i_s}^{ j_1 \cdots j_s} (\k)
\equiv \sum_{\lambda} \epsilon^\lambda _{i_1 \cdots i_s} (\k) \epsilon^{* j_1 \cdots j_s}_{\lambda } (\k).
\ee
It was also assumed that the background generated by the IR modes breaks the isotropy by identifying a constant unit vector $\hat{p}_{i}$ ($\hat{p} \cdot \hat{p} = 1$) such that
\begin{eqnarray}
&& \Braket{\bar{A}_{i_1 \cdots i_s}} = \bar{A_0} \left[\hat{p}_{i_1}\cdots \hat{p}_{i_s} \right. \nonumber \\
  && \qquad \left. -\frac{1}{2s-1} \left(\delta _{i_1 i_2} \hat{p}_{i_3} \cdots \hat{p}_{i_s} +  {\rm perms} \right)+ \cdots \right], \label{eq:vac}
\end{eqnarray}
where the ellipsis stands for further terms ensuring the r.h.s. is transverse and traceless.
Then, the angle-averaged bispectrum (averaging over $\hat p$), was found to be
\begin{eqnarray} 
  B_\zeta(k_1,k_2,k_3) 
  &=& \frac{1}{2} g_s I^0_s P_\zeta(k_1)P_\zeta(k_2) \nonumber \\ 
  &&  \hspace{-10pt} \times  \left\{(1+\cos\theta_{\hat{k}_1, \hat{k}_2})^{2s}
 +(1-\cos\theta_{\hat{k}_1, \hat{k}_2})^{2s}\right\} \nonumber \\ 
   && + 2~{\rm perms}.
\end{eqnarray}
where we have defined $I^0_s=\bar{A}_0 ^2 s!/(2s+1)!!$ and $g_s$ is an undetermined constant. 
We expect  $g_s$ to be of order unity and $\bar{A_0}$ to be of order of  $H N^{1/2}$ where $H$ is the Hubble rate and $N$ is the total number of
e-folds of inflation.  The observational limits discussed in this paper can therefore provide  useful informations about the number of e-folds if $H$ is known by some alternative method.
Finally, the coefficients $C_n$ appearing in Eq. \eqref{eq:temp} are 
\begin{equation}
    \begin{aligned}
      C_0  & = \frac{4^{s} g_s I_s^0 }{(2s + 1 )} , \\
      C_n  & = \frac{ C_0 (2 n+1) \Gamma (2 s+1) \Gamma (2 s+2)}{\Gamma
   (-n+2 s+1) \Gamma (n+2 s+2)} \ \ \ [n = {\rm even}], \\
      C_n &= 0 \ \ \ [n > 2 s \ \text{or} \ n = {\rm odd}].
    \end{aligned}
  \label{eq:cn_theory}
\end{equation}
It is interesting to note that the same combination of undetermined constants $g_s \bar{A}_0^2$ appears in all even $C_n$.

In an analogous way, the angle-averaged bispectrum contribution resulting from partially massless spinning particles takes the form \cite{Franciolini:2017ktv}:
\begin{equation}
  B_\zeta(k_1,k_2,k_3)  = g_s^{\text{PM}} I^0_s  P_\zeta(k_1)P_\zeta(k_2) \Pi_s+  2~{\rm perms},
\label{eq:bis-pm}
\end{equation}
where 
\begin{equation}
  \Pi_s = \sum_{\lambda_1 \lambda_2\neq 0,\pm1}  d_{s|\lambda_1|}d_{s|\lambda_2|} 
  \left| \epsilon_{\lambda_1, i_1 \cdots i_s}(\k_1) \epsilon^*_{\lambda_2, i_1 \cdots i_s}(\k_2) \right|^2\!.
\end{equation}
The constants $d_{s|\lambda|}$, normalised so that $d_{ss}=1$, account for the mixing of different helicities of the partially massless field due to the different expectation values of its $s-2$ components.    As such, they should be treated as free and independent parameters since the expectation values are  not predicted  by the underlying theory.
For the simplest case $s=3$
 the coefficients $C_n$ in Eq.~\eqref{eq:temp} are 
 \begin{equation}
  \begin{aligned}
    C_{0} &= \frac{16}{7}(2+3 d_{32})^2g_s^{\text{PM}} I^0_s,  \\
    C_2 &= \frac{400}{21} g_s^{\text{PM}} I^0_s , \\
    C_4 &= \frac{1}{66} \left(7  \sqrt{3 C_0}+12
   \sqrt{C_2}\right)^2,  \\
    C_6 &= \frac{1}{275} \left(5 \sqrt{3 C_0}+7
   \sqrt{C_2}\right)^2,  \\
    C_n &=0 \ \ \ [n \neq 0,2,4,6] .
  \end{aligned}
\label{eq:cn_theory_1}
\end{equation}
Analogously, for $s=4$, it was found that
\begin{equation}
 \begin{aligned}
C_{0} =& \frac{256}{315} \Big\{972 d_{42}^2 + 360 d_{42} (1 + d_{43}) \\
&\qquad  +35 (1 + 2 d_{43} + 18 d_{43}^2)\Big\}g_s^{\text{PM}} I^0_s, \\
C_2 =& \frac{512}{693}    \Big\{648 d_{42}^2 + 49 (2 + d_{43})  \\
&\qquad - 72 d_{42} (4 + d_{43})\Big\}  g_s^{\text{PM}} I^0_s , \\
C_4 =& \frac{512}{5005} \Big\{245 + 11763 d_{42}^2 - 735 d_{43} \\
&\qquad + 990 d_{42} (-2 + 3 d_{43})\Big\}g_s^{\text{PM}} I^0_s,  \\
C_6 =& \frac{512}{3465}\Big\{14 + 11664 d_{42}^2 - 119 d_{43}  \\
&\qquad - 198 d_{42} (-4 + 17 d_{43})\Big\}g_s^{\text{PM}} I^0_s,  \\
C_8 =& \frac{128}{6435}(1 + 144 d_{42}) \\
& (1 + 144 d_{42} - 16 d_{43})  g_s^{\text{PM}} I^0_s  ,  \\
C_n =& 0 \ \ \ [n \neq 0,2,4,6,8].
    \end{aligned}
\label{eq:cn_theory_2}
\end{equation}
Note that in this case, in addition to the combination $g_s \bar{A}_0^2$, certain combinations of $d_{s|\lambda|}$ appear. 

Note that the template of the form given in Eq. \eqref{eq:temp} has been initially introduced in Ref. \cite{Shiraishi:2013vja} to characterize the angle-averaged bispectrum sourced by the presence of a $U(1)$ gauge vector field coupled to inflaton field via the interaction $I(\phi)F^2$ during inflation \cite{Barnaby:2012tk,Bartolo:2012sd}, models with primordial magnetic fields  \cite{Shiraishi:2012rm}, and models with non-trivial symmetry structure of inflaton field as in solid inflation \cite{Endlich:2012pz,Endlich:2013jia}. These models can generate angular dependence corresponding to the first three terms in the above expansion. Therefore, this template is used to forecast the constraints on $C_{0,1,2}$, from the CMB and LSS (see for ex. \cite{Shiraishi:2013vja,Raccanelli:2015oma, Schmidt:2015xka,Kogai:2018nse}). Moreover, observational bounds for $C_{0,1,2}$ are obtained via the measurements of the CMB bispectra by the Planck satellite \cite{Ade:2015ava}. In Ref. \cite{Franciolini:2017ktv,Franciolini:2018eno}, the higher-order (even) terms in the expansion are considered. They performed a forecast to obtain constraints on $C_{0,2,4,6,10}$ from measurement of CMB temperature bispectrum \cite{Franciolini:2018eno}. We extend their work by making a forecast for these coefficients using the observed galaxy bispectrum from upcoming galaxy surveys.

\section{The observed galaxy bispectrum}\label{sec:gbis}
Our model of the galaxy bispectrum is the same as Ref. \cite{MoradinezhadDizgah:2018ssw}, so we would not repeat the details here and refer the reader to that reference. To summarize, we model the galaxy bispectrum at tree-level in perturbation theory accounting for redshift-space distortions (RSD) (linear Kaiser term and Finger-of-God effect) in addition to Alcock-Paczynski (AP) effect. In relating the galaxy overdenisty to that of underlying dark matter, we assume a simple model of the bias, accounting for local-in-matter terms up to quadratic order as well as the tidal shear bias. 
 
Neglecting the AP effect, let us briefly review the model of galaxy bispectrum we use in our forecast. In addition to the contribution from the non-vanishing primordial bispectrum, non-linear gravitational evolution generates non-zero bispectrum of matter density field and biased tracers.  At leading-order in perturbation theory, the total bispectrum at redshift $z$ is the sum of the two contributions
\begin{align}
B_g&(\k_1,\k_2,\k_3;z) \nonumber \\
&\equiv B_g^{\rm grav}(\k_1,\k_2,\k_3)  + B_g^{\rm PNG}(\k_1,\k_2,\k_3)\,,
\end{align}
where the contribution from gravitational evolution is  
\begin{align}
B_g^{\rm grav}\!(\k_1,&\k_2,\k_3) = D_{\rm FoG}^B(\k_1,\k_2,\k_3)\left[2Z_1(\k_1)Z_1(\k_2) \right. \nonumber \\
&\times \left. Z_2(\k_1,\k_2)P_0(k_1)P_0(k_2) + \text{perms}\right]\,, 
\end{align}
while the contribution from primordial bispectrum $B_\zeta$ is given by
\begin{align}
B_g^{\rm PNG}(\k_1,&\k_2,\k_3) = D_{\rm FoG}^B(\k_1,\k_2,\k_3)\nonumber  \\
&\times \prod_{i=1}^3\left[ Z_1(\k_i){\mathcal M}(k_i) \right] B_\zeta(k_1,k_2,k_3)\,. 
\end{align}
The kernels $Z_i$ are the perturbation theory kernels in redshift space and $D_{\rm FoG}^B$ is the finger-of-God suppression factor. $P_0$ denotes the matter power spectrum linearly extrapolated to redshift $z$, and ${\cal M}(k)$ is the transfer function that relates the primordial fluctuations $\zeta$ to the linearly extrapolated matter overdensity during the matter-domination era. The explicit expressions of these functions are given in Ref. \cite{MoradinezhadDizgah:2018ssw}. Note that for brevity,  we have dropped the explicit redshift dependence in the above expressions.  Momentum conservation $\k_1+\k_2+\k_3 = 0$ removes the dependence on one of the wave-vectors. Even if the primordial bispectrum is isotropic, RSD and AP effects, render the observed bispectrum anisotropic. Therefore to characterize galaxy bispectrum in addition to the shape of the triangle, one needs to define the orientation of the triangle with respect to the line of sight. Hence, the bispectrum depends on five independent parameters which can be chosen to be the three sides of the triangle $k_1,k_2,k_3$, the angle $\theta$ between one of the wavevectors and the line-of-sight,  and the azimuthal angle $\phi$ between two wavevectors.

\section{Forecasting Methodology}\label{sec:forecast_meth}
We use the Fisher matrix formalism to study the potential of the upcoming EUCLID spectroscopic survey, in setting constraints on the presence of long-lived particles with spin. Our forecasting methodology and the survey specifications are the same as Ref.\cite{MoradinezhadDizgah:2018ssw}, so here we only briefly review our assumptions and refer the reader to this reference for details.  

We use the survey specifications of EUCLID spectroscopic survey as outlined in \cite{Laureijs:2011gra,Amendola:2016saw}. We assume a sky coverage of 15000 ${\rm deg}^2$, i.e. $f_{\rm sky} = 0.36$, and redshift uncertainty of $\sigma_z(z) = 0.001(1+z)$. The redshift distribution $dN/dz$ is obtained from empirical data of luminosity function of H$\alpha$ emitters, and we take the limiting flux of $4\times 10^{-16} {\rm erg} s^{-1} {\rm cm}^{-2}$ and efficiency of $35\%$. We consider 12 equally populated redshift bins in the range of $0.4<z<2.1$, similar to what is done in Ref. \cite{Giannantonio:2011ya}.

The Fisher matrix of the galaxy bispectrum at a given redshift bin with mean $z_i$ is given by
\begin{align}
	F_{\alpha \beta}(z_i) &= \frac{V_i}{(2\pi)^5} \int_{{\mathcal V}_B}  d V_k\, k_1 k_2 k_3\int_{-1}^1  d \cos \theta \int_0^{2\pi}\d \phi\, \nonumber \\
	&\times \frac{(\partial B_g^{\rm obs}/\partial\lambda_\alpha)(\partial B_g^{\rm obs}/\partial\lambda_\beta)}{{\rm Var} B_g}\, ,
\label{eq:FisherG}
\end{align}
where we defined $dV_k \equiv dk_1 dk_2 dk_3$, ${\mathcal V}_B$ is the tetrahedral domain allowed by triangle condition for the wavenumbers $k_{\rm min} < k_i<k_{\rm max}$, and $V_i$ is the volume of the redshift bin $z_i$. For each redshift bin, we take $k_{\rm min} = 2\pi(3V_i/4\pi)^{-1/3}$ and set $k_{\rm max} $ such that the variance of the matter density field at that redshift is equal to the variance at $z=0$ for $k_{\rm max} =0.15 \ h\ {\rm Mpc}^{-1}$. We also impose a conservative upper bound that $k_{\rm max} \leq 0.3 \ h\ {\rm Mpc}^{-1}$. The total Fisher matrix is the sum of the Fisher matrices in all redshift bins covered by the survey 
\be
F_{\alpha \beta} = \sum_i F_{\alpha \beta}(z_i)\, .
\ee
For the variance of the bispectrum,  in our main analysis, we only consider the Gaussian contribution which is given by
\begin{align}
{\rm Var} \ &B_g(k_1,k_2,k_3,\theta,\phi,z_i) \nonumber \\
&= s_{123} \prod_{j=1}^3\left[P_g^{\rm obs}(k_j,\mu_j,z_i) + \frac{1}{\bar n_i}\right],
\end{align}
where $P_g^{\rm obs}$ is the observed power spectrum, with $\mu_j$ being the angle of each wavevector with the line of sight. $\bar n_i$ is the mean number density of galaxies in redshift bin $z_i$ and $s_{123} = 6,2,1$ for equilateral, isosceles, and scalene triangles. The constraints we report in section \ref{sec:results} are obtained imposing Planck priors, as discussed in Ref. \cite{MoradinezhadDizgah:2018ssw}, using synthetic temperature data, therefore
\be
F_{\alpha \beta}^{\rm tot} = F_{\alpha \beta} + F_{\alpha \beta}^{\rm Planck}\,.
\ee

In our forecast we consider three cases: only one of the coefficients $C_n$ is non-zero, all even $C_n$ coefficients up to $n\leq 10$ are non-zero and the coefficients are all a function of $C_0$. In addition to the coefficients $C_n$, we vary five cosmological parameters; amplitude ${\rm ln} (10^{10}A_s)$, and spectral index $n_s$ of primordial scalar fluctuations, Hubble parameter $h$, and the energy density of cold dark matter $\Omega_{\rm cdm}$, and that of baryons $\Omega_b$. We also vary three biases, linear and quadratic local biases $b_1,b_2$ and tidal shear bias $b_{K^2}$. For the biases, we assume that the redshift evolution is known (as discussed below), and we vary a single parameter characterizingThe suppression factor of FoG effect is determined by dispersion velocity of galaxies, for which we assume that the redshift evolution is known and vary a single parameter $\sigma_{{\rm FOG},0}$.  In obtaining the constraint on each coefficient $C_n$, our parameter array is therefore ${\boldsymbol\lambda^{(i)}} = \left[{\rm ln} (10^{10}A_s), n_s, h,\Omega_{\rm cdm},\Omega_b, C_n,\sigma_{{\rm FOG},0}, b_1, b_2, b_{K^2} \right]$. 

We choose the fiducial values of the cosmological parameters to be $\ln(10^{10}A_s) = 3.067, n_s = 0.967, h = 0.677$, $\Omega_{\rm cdm} =0.258$, $\Omega_b = 0.048$, setting the pivot scale of $k_p=0.05 \ {\rm Mpc}^{-1}$, consistent with the Planck 2015 data \cite{Ade:2015xua}. For the $C_n$ coefficients we set the fiducial values of $C_n = 1$, and for the velocity dispersion we set the fiducial value to be $\sigma_{{\rm FOG},0} = 250 \ {\rm km }$ s$^{-1}$ (similar to Ref. \cite{Giannantonio:2011ya}). We model the redshift evolution of the linear bias as $b_1(z) = \bar b_1 \sqrt{1+z}$ and set the fiducial value of $\bar b_1 = 1.46$ such that at $z=0$ the value of the linear bias is consistent with the results of Ref. \cite{Lazeyras:2015lgp} for halos of mass $M = 3\times 10^{13} h^{-1}M_\odot$. For quadratic biases we assume scaling relations of $b_2 = \bar b_2(0.412 -  2.143 b_1 + 0.929 b_1^2 + 0.008 b_1^3)$ and $b_{K^2} = \bar b_{K^2}(0.64 - 0.3 b_1 + 0.05 b_1^2 -0.06 b_1^3)$, which are fits to N-body simulations provided in Refs.~\cite{Lazeyras:2015lgp,Modi:2016dah}. Based on these results, we assume that the above relations between $b_2$ and $b_{K^2}$ with $b_1$, are preserved in the redshift range we consider and use it to set the fiducial values of the the biases in each redshift bin. We vary two parameters for the overall amplitudes $\bar b_2$ and $\bar b_{K^2}$.

Additionally we will also consider how the forecasted constraints on the coefficients $C_n$ degrade, once the uncertainty in the theoretical model of the galaxy bispectrum \cite{Baldauf:2016sjb,Karagiannis:2018jdt} and leading non-Gaussian corrections to the variance \cite{Chan:2016ehg,Karagiannis:2018jdt} are accounted for. We will follow the same prescriptions as reviewed in \cite{MoradinezhadDizgah:2018ssw}, therefore we refer the reader to Ref. \cite{MoradinezhadDizgah:2018ssw} and references therein for the details. 

\section{Results}\label{sec:results}
As discussed in section \ref{sec:theory},  both in the case of a HS field coupled to inflaton and that of the partially massless HS field, different $C_n$ coefficients are related to one another. In our analysis, being agnostic to the theoretical model, we first consider the case where the coefficients $C_n$ are independent of one another.  We obtain constraints on $C_{n_{0,2,...,10}}$, assuming only one is non-zero (shown in Table \ref{tab:res_sep}), or all are non-zero (shown in table \ref{tab:res_all}).  Next we consider the case of the field with spin $s$ coupled to inflaton, where all the even $C_{n\leq 2s}$ coefficients can be written in terms of $C_0$, and obtain constraints on $C_0$ for a given spin (shown in table \ref{tab:res_C0}). In all the tables below, the three columns correspond to using the Gaussian expression for the variance (``Base"), accounting for the leading non-Gaussian correction to the variance (``NG Var.''), and accounting for the theoretical error (``TH Err.''). The constraints are obtained marginalizing over cosmological parameters, biases and the dispersion velocity as described in section \ref{sec:forecast_meth}. 

 In Table \ref{tab:res_sep}, we show the 1-$\sigma$ constraints on $C_{0,2,...,10}$, varying one at the time and assuming all the others are zero. As a consistency check of our forecasting pipeline, note that for $n=0$, the template in Eq. \eqref{eq:temp} reduces to the local shape with $C_0 = 6/5 f_{\rm NL}^{\rm loc}$. Therefore the constraint for $C_0$ is in agreement with that of Ref. \cite{MoradinezhadDizgah:2018ssw} for the local shape.  The constraints get weaker as $n$ increases. Accounting for the leading NG correction to the variance degrades the constraints by about a factor of $(30-40)\%$, while taking into account the theoretical error weakens the constraints by less than $10\%$.
\renewcommand{\arraystretch}{2}
\begin{table}
\begin{tabular}{| c | c | c | c| }
\hline 
  & \  \  \ ${\rm Base}$ \  \ & \  ${\rm NG \ Var.}$ \  &  \  ${\rm TH \ Err.}$ \ \\ 
\hline \hline
\ \ $\sigma(C_0)$       \  \  &  0.451  &  0.610   & 0.490  \\ \hline
\ \ $\sigma(C_2)$       \  \  &  0.895 &  1.22  & 0.981 \\ \hline
 \ \ $\sigma(C_4)$      \  \  &  1.03   &  1.40  &  1.13 \\ \hline
  \ \ $ \sigma(C_6)$    \  \  &  1.24   &  1.61  & 1.29 \\ \hline
  \ \ $\sigma(C_8)$     \  \  &  1.43  &  	1.96  & 1.55 \\ \hline
  \ \ $\sigma(C_{10})$ \  \ &  1.58   &  2.17  & 1.71 \\ \hline
\end{tabular}
\caption{1-$\sigma$ constraints on the coefficients $C_n$, varying one at the time and setting the rest to zero. The constraints are obtained marginalizing over cosmological parameters and biases. We chose $C_n =1$ as our fiducial values. The fiducial values of biases and cosmological parameters are given in the text. }
\label{tab:res_sep} 
\end{table} 

The current best constraint on the lowest order coefficients are obtained from measurement of CMB temperature and polarization bispectra by Planck satellite which provided $\sigma(C_0) =6 $ and $\sigma(C_2) = 26$ \cite{Ade:2015ava}. For higher-order terms, comparing our forecasted constraints with those for CMB temperature bispectrum in \cite{Franciolini:2018eno}, our result indicate that the measurement of galaxy bispectrum would provide significantly tighter constraints, which can be attributed to having access to more modes since LSS is a 3-dimensional map of the universe in contrast with CMB which is a 2-dimensional map.  Moreover, unlike the constraints from CMB \cite{Franciolini:2018eno}, in which the constraints on higher-order coefficients are significantly weaker than lowest order ones, the LSS constraints on coefficients with $n=0, ..., 10$ are comparable.  We believe this can be understood in the following way: the CMB bispectrum probes the projected primordial bispectrum in two dimension, hence, the oscillatory features of the bispectrum is washed away. The impact is more important for higher order Legendre polynomials since they are highly oscillatory. 

\renewcommand{\arraystretch}{2}
\begin{table}[htbp!]\centering
\begin{tabular}{| c | c | c | c| }
\hline 
  & \  \  \ \ \  ${\rm Base}$ \  \  \ \ & \  ${\rm NG \ Var.}$ \  &  \  ${\rm TH \ Err.}$ \ \\ 
\hline \hline
\ \ $\sigma(C_0)$       \  \  &  1.23 (0.275) & 1.72  &  1.45\\ \hline
\ \ $\sigma(C_2)$       \  \  &  4.28 (0.688) &  6.22 &  5.08\\ 	\hline
 \ \ $\sigma(C_4)$      \  \  &  6.22 (0.930) &  9.12 & 7.25 \\ \hline
  \ \ $ \sigma(C_6)$    \  \  & 7.05  (1.12) & 10.3 	&  8.21 \\ \hline
  \ \ $\sigma(C_8)$     \  \  &  6.51 (1.35) & 9.35 	&  7.59 \\ \hline
  \ \ $\sigma(C_{10})$ \  \ &  4.20  (1.50) & 5.88 	& 4.85 \\ \hline
\end{tabular}
\caption{1-$\sigma$ constraints on each $C_n$ when all varied, marginalizing over the other coefficients as well as cosmological parameters and biases. The fiducial values of biases and cosmological parameters are given in the text. The numbers in the parentheses are the un-marginalized constraints. }
\label{tab:res_all} 
\end{table} 
\begin{figure}\centering
\hspace{-.3in}\includegraphics[width=0.45\textwidth]{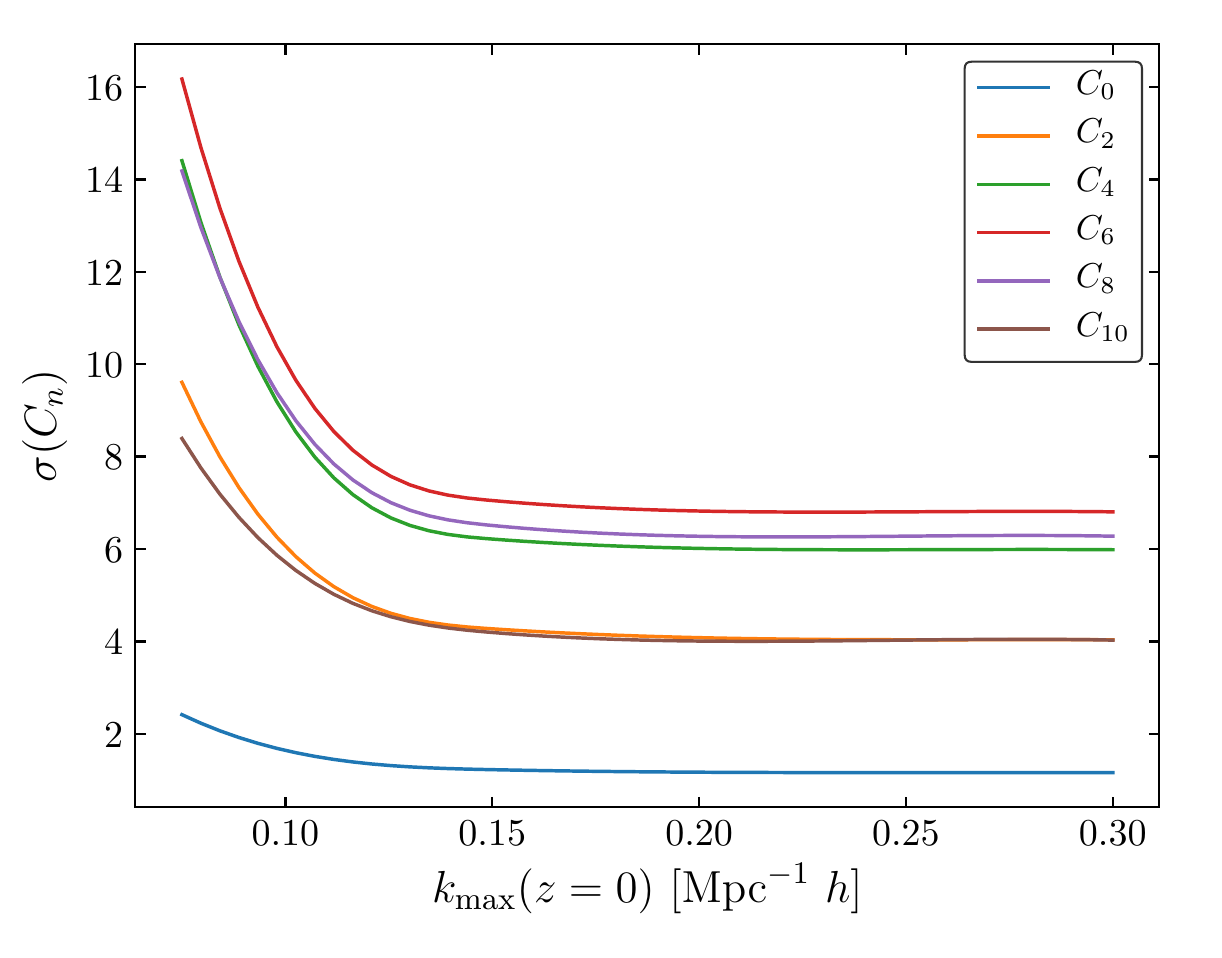}
\caption{1-$\sigma$ confidence ellipses for the coefficients $C_n$ from EUCLID survey as a function of $k_{\rm max}(z=0)$. The fiducial values of biases and cosmological parameters are given in the text.}
\label{fig:Cn_kmax}
\end{figure}

\begin{figure}
\hspace{-.3in}\includegraphics[width=0.524 \textwidth ]{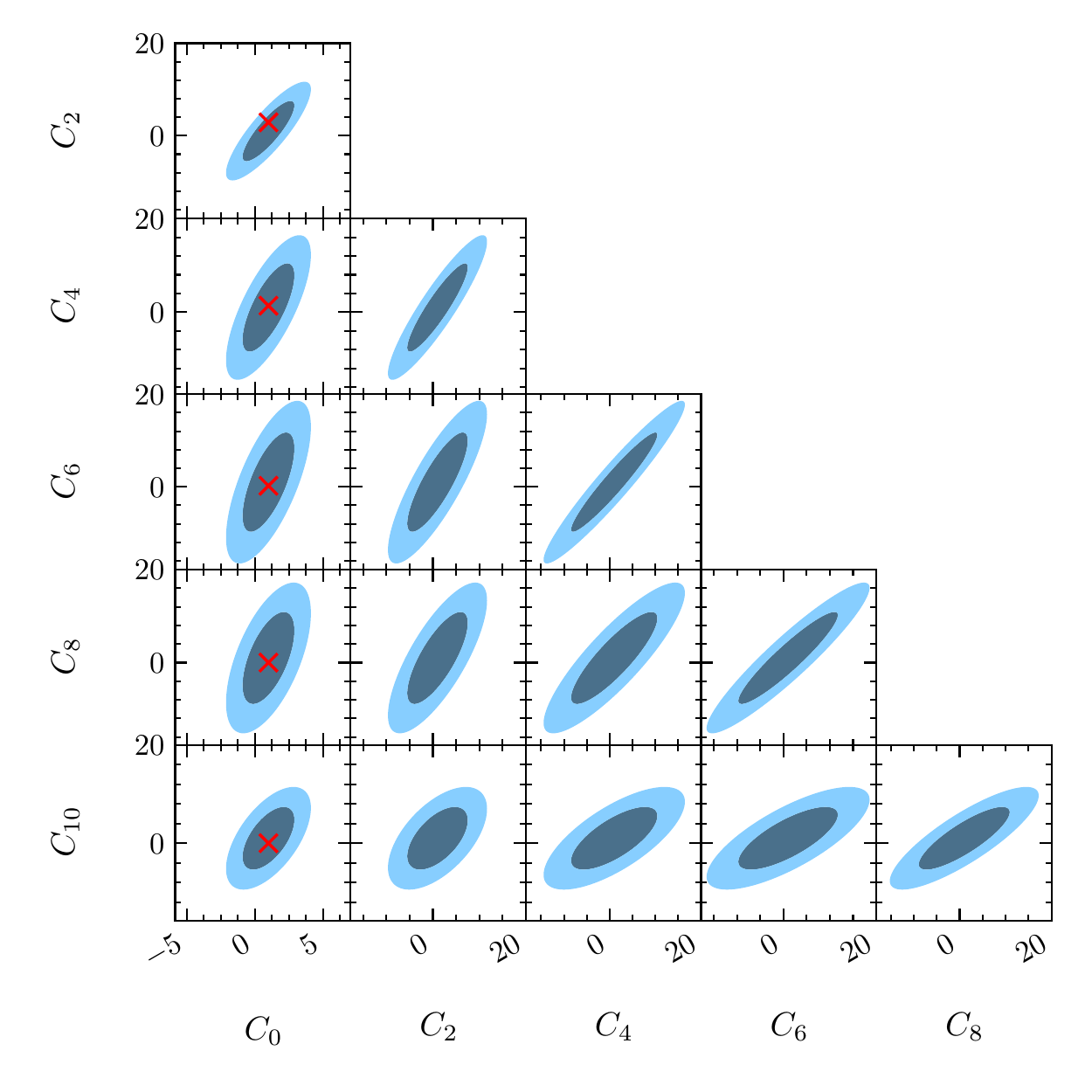}
\caption{1- and 2-$\sigma$ confidence ellipses for the coefficients $C_n$ from EUCLID survey, marginalizing over all the other parameters. We chose $C_n =1$ as our fiducial values. The fiducial values of biases and cosmological parameters are given in the text.}
\label{fig:Cns}
\end{figure}
\begin{figure}
\centering
\hspace{-.3in}\includegraphics[width=0.524 \textwidth ]{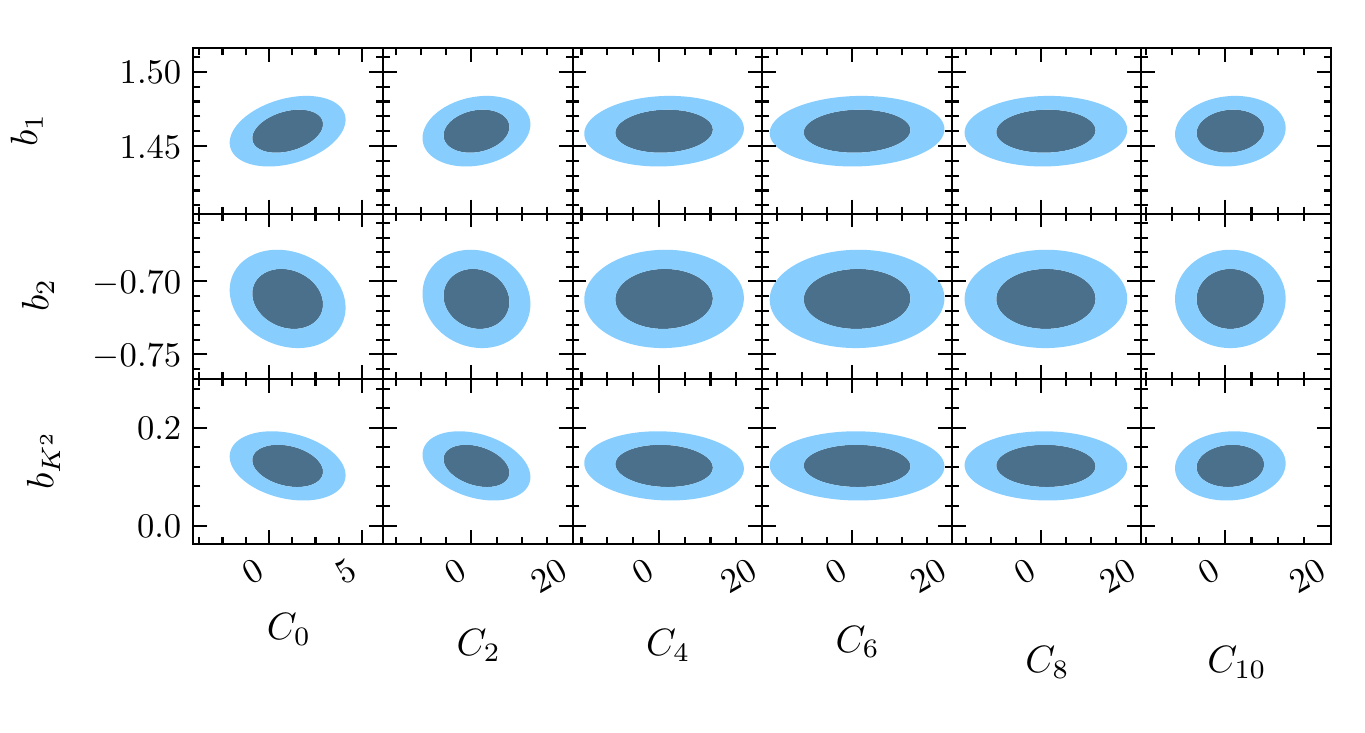}
\caption{1- and 2-$\sigma$ confidence ellipses for the biases and the coefficients $C_n$ from EUCLID survey, marginalizing over all the other parameters. The fiducial values of biases and cosmological parameters are given in the text.}
\label{fig:bias_Cn}
\end{figure}

Since the theoretical models considered here, predict a subset of the $C_n$ coefficients to be non-zero for a given spin, to study the degeneracy between various $C_n$ coefficients, next we assume that $C_{0,2,..10}$ are non-zero simultaneously. We obtain the 1-$\sigma$ constraints on each of the $C_n$, marginalizing over the others, as well as cosmological parameters, biases and dispersion velocity.  The results are shown in Table \ref{tab:res_all}. Overall, the constraints on each coefficient are weaker than in the case of varying only one at a time, due to degeneracy between them. Among all,  the  constraints on the coefficient $C_6$ are the weakest and improves for $n>6$. We note that if considering the un-marginalized constraints (the numbers in the parentheses), the constraints degrade for increasing $n$. Therefore the improvement in the marginalized constraints for $n>6$ is due to parameter degeneracies. Accounting for the leading non-Gaussian correction to the variance and the theoretical error, degrade the constraints by about a factor of $(40-50)\%$ and $(15-20)\%$, respectively. To show the dependence of the constraints on the choice of $k_{\rm max}$, in Fig. \ref{fig:Cn_kmax}, we show the $1-\sigma$ constraints on $C_n$ as a function of $k_{\rm max}$ at redshift $z=0$. Note that plateauing of the constraint is partially due to the fact that we always impose the upper bound of $k_{\rm max} \leq 0.3 \ h \ {\rm Mpc}^{-1}$. Figures \ref{fig:Cns} and \ref{fig:bias_Cn} show the 1- and 2-$\sigma$ confidence ellipses between $C_n$ coefficients and between $C_n$ coefficients and the three biases, respectively, marginalizing over all the other parameters.  There is a significant degeneracy between various $C_n$ coefficients  and a non-negligible degeneracy between $C_0$ and $C_2$ with the biases as shown in Fig. \ref{fig:bias_Cn}. The degeneracy between the coefficients $C_n$ and cosmological parameters is rather week, and hence, we do not show it here. To compare with the prediction of the theoretical models described in section \ref{sec:theory}, in Fig. \ref{fig:Cns}, as an example we also show the values of $C_n$ for $n \neq 0$ in terms of $C_0$ for the case of a particle with spin $s=5$ coupled to inflaton as given in Eq. \eqref{eq:cn_theory}. 

Next, we consider the case of the HS field coupled to inflaton, in which the coefficients $C_n$ are all related to $C_0$ as give in Eq. \eqref{eq:cn_theory}. For a given spin $s$, only even coefficients up to $n \leq 2s$ are non-zero. We consider particles with spins $1,2,3,4,5$ and obtain the constraint on $C_0$ in each case, marginalizing over all the other parameters. The results are given in Table \ref{tab:res_C0}. 
\renewcommand{\arraystretch}{2}
\begin{table}
\begin{tabular}{| c | c | c | c| }
\hline 
  & \  \  \ ${\rm Base}$ \  \ & \  ${\rm NG \ Var.}$ \  &  \  ${\rm TH \ Err.}$ \ \\ 
\hline \hline
\ \ $s=1$    \  \   &  0.301 &  0.439  & 0.328 \\ \hline
\ \  $s=2$    \  \  &  0.189 & 0.246   & 0.199 \\ \hline
 \ \  $s=3$   \  \  &  0.138 & 0.176   & 0.143 \\ \hline
  \ \  $s=4$  \  \  & 0.110  &  0.138  &  0.114 \\ \hline
  \ \  $s=5$  \  \  & 0.093  & 0.114   & 0.095\\ \hline
\end{tabular}
\caption{1-$\sigma$ constraints on $C_0$ from massive particles with spins $s=1,2,3,4,5$ coupled to inflaton. Constraints are obtained marginalizing over cosmological parameters and biases.}
\label{tab:res_C0} 
\end{table} 
In this case, as one would expect, we obtain better constraints on $C_0$ than previously, since there are additional contributions from higher order Legendre polynomials to $C_0$. The constraints are degraded by about a factor of $(20-40)\%$, when we account for the leading NG correction to the variance and by less than $10\%$ when theoretical error is accounted for. Note that if we were to measure $C_n$ of various orders, a non-zero value of $C_0$ and zero value for all higher-order terms would correspond to detecting the local shape non-Gaussianity. However if we only measure $C_0$, we can not infer that, the signal is due to the local shape unless we measure other coefficients to be zero. 

\begin{figure}[htbp!]
\hspace{-.32in}\includegraphics[width=0.528 \textwidth ]{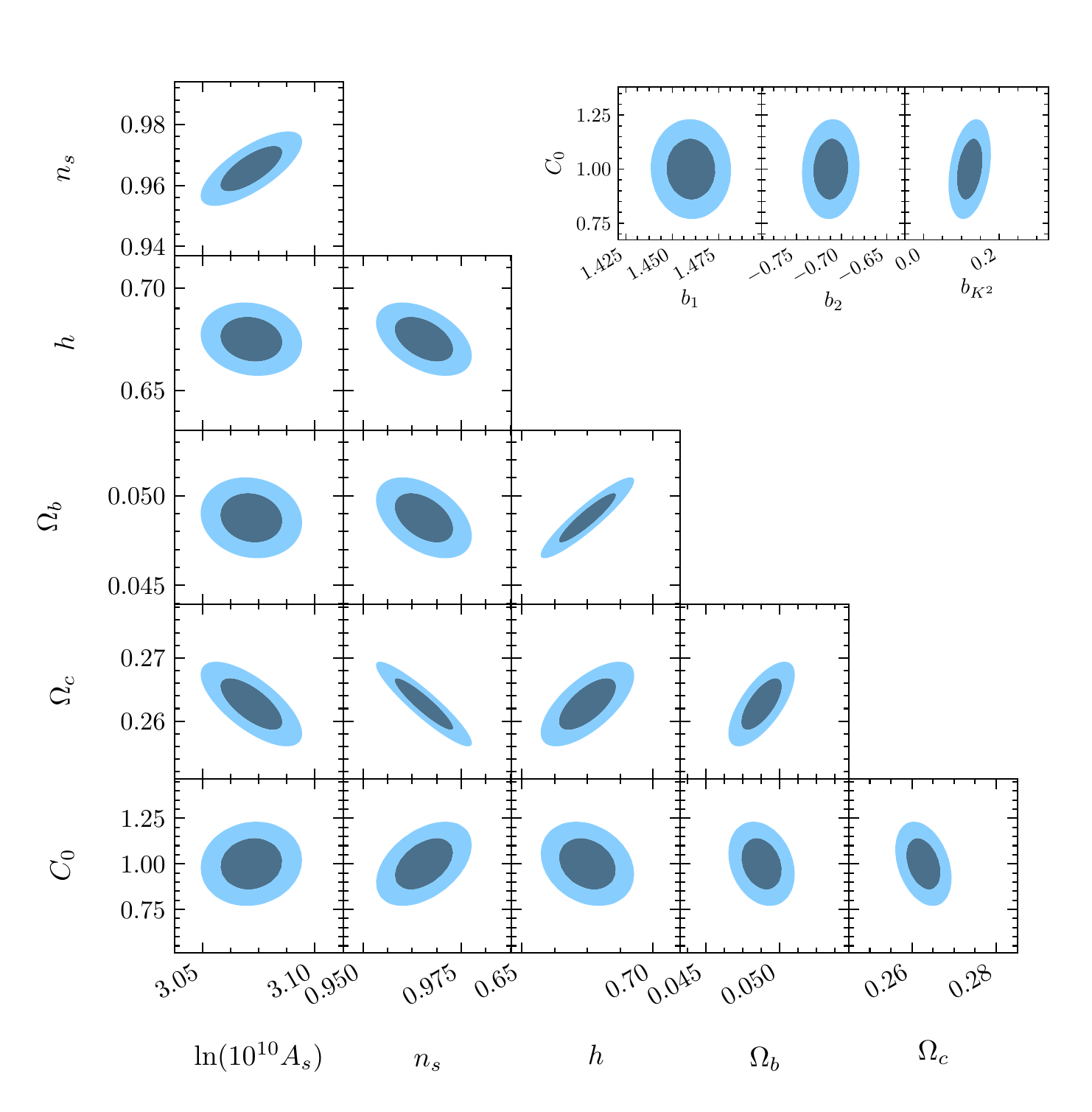}
\caption{1- and 2-$\sigma$ confidence ellipses for $C_0$ from HS massive particles with $s=5$, and cosmological parameters and the biases from EUCLID survey, marginalizing over all the other parameters. The fiducial values of biases and cosmological parameters are given in the text.}
\label{fig:C0_all}
\end{figure}

To demonstrate the degeneracy between $C_0$ and other parameters, for the case of a HS field with $s=5$, we show the 1- and 2-$\sigma$ confidence ellipses between $C_0$ and the cosmological parameters, as well as with the  biases in Fig. \ref{fig:C0_all}. The constraints are obtained marginalizing over all the other parameters. 

\section{Conclusions}\label{sec:conclusions}
Particles with non-zero spin, if present during inflation, leave a distinct angular dependance on correlation functions of primordial curvature fluctuations. For massive particles, the amplitude of the signal, in particular the bispectrum, of primordial scalar fluctuations, however, is in general suppressed with suppression factor determined by the mass of the particles. Recently it has been shown that by introducing a suitable coupling between the particles with spin and inflaton, or by considering the partially massless particles, one can generate long-lived particles that lead to  un-suppressed primordial bispectrum. The resulting bispectrum in these models are shown to be anisotropic as the non-zero vacuum expectation value of the particle with spin, introduces a preferred direction. After angular averaging this bispectrum, one can express the results as a finite expansion in terms of Legendre polynomials. The contributing terms and the relation between the coefficients of the expansion are fully determined in each model. 

In this paper we investigated the potential of the upcoming EUCLID spectroscopic survey in constraining the coefficients of this expansion, and hence in setting constraints on the presence of long-lived extra particles with spin, described by these models.  Assuming that the coefficients of this expansion are independent of one another, we considered terms up to 10-th order and showed that measurement of the galaxy bispectrum from the EUCLID survey can potentially constrain them with an uncertainty of order unity. We additionally considered the case where the relation between the coefficients is determined by the theoretical model (the HS field coupled to inflaton). In this case the primordial bispectrum receives contribution from all the even terms in the expansion with $n\leq 2s$ and the amplitude of all is proportional to the zeroth-order term $C_0$. Therefore, in this case the constraints on $C_0$ improve  by about an order of magnitude. Furthermore, within the assumptions of our prescriptions, we showed that the non-Gaussian contribution to the variance has a more significant impact on the constraints than the uncertainty in theoretical modeling of the observed galaxy bispectrum (neglecting the higher-order loops). 

As a last comment, let us reiterate  that, if in the future a local non-Gaussian linear parameter (corresponding to $C_0$) will be measured to be significantly different from zero, 
one should make an effort to detect the higher-multipoles of the bispectrum in order to be sure that one is not dealing with a higher-spin field. Conversely, detecting the higher-multipoles will be an indication that higher-spin fields play a role during inflation.

\begin{acknowledgments}
It is our pleasure to thank Julian Mu\~noz for helpful discussions. A.R. is supported by the Swiss National Science Foundation (SNSF), project {\sl Investigating the Nature of Dark Matter}, project number: 200020-159223. 

\end{acknowledgments}

\bibliography{SCINC}

\end{document}